# A Network Filtration Protocol for Elucidating Relationships between Families in a Protein Similarity Network

Leonard Apeltsin

## Abstract

**Motivation:** The study of diverse enzyme superfamilies can provide important insight into the relationships between protein sequence, structure and function. It is often challenging, however, to discover these relationships across a large and diverse superfamily. Contemporary similarity network visualization techniques allow researchers to aggregate sequence similarity information into a single global view. Network visualization provides a qualitative estimate of functional diversity within a superfamily, but is unable to quantitate explicit boundaries, when present, between neighboring families in sequence space. This limits the potential of existing sequence-based algorithms to generate functional predictions from superfamily datasets.

**Results:** By building on current network analysis tools, we have developed a new algorithm for elucidating pairs of homologous families within a sequence dataset. Our algorithm is able to filter through a dense similarity network in order to estimate both the boundaries of individual families and also how the families neighbor one another. Globally, these neighboring families define a topology across the entire superfamily. The topology is simple to interpret by visualizing the network output generated by our filtration protocol. We have compared the network topology within the kinase superfamily against available phylogenetic data. Our results suggest that neighbors within the filtered kinase network are more likely to share structural and functional properties than more distant network clusters.

## 1 Introduction

Some homologous but highly divergent sets of proteins have evolved to perform substantially different molecular functions. These include a wide range of membrane transporters (George et al., 2004) as well as mechanistically diverse enzyme superfamilies (Pegg et al., 2006). Mechanistically diverse enzyme superfamilies are sets of evolutionarily related proteins with similar structural and functional properties. All members of such superfamilies share the same structural scaffold and use a conserved subset of active site residues that can be associated with an underlying aspect of catalysis, often a partial reaction (Babbitt and Gerlt, 1997), (Babbitt and Gerlt, 2001). A superfamily can further be subdivided into individual families. Each family catalyzes a unique overall reaction which, together with a distinct set of catalytic residues, differentiates it from all the other families in the set (Pegg et al., 2005). Each individual family within a superfamily can usually be further differentiated by its substrates and products.

Given a superfamily with a few hundred or more protein sequences, it would be valuable to summarize how families within the superfamily relate to one another. More specifically, we would like to extract individual families from the dataset and determine which pairs of families share the strongest degree of functional similarity with each other. We restrict ourselves here only to sequence information because it is widely available and provides us access to large amounts of data. Aggregating these pairs of "neighboring" families allows us to define a topology that in some cases can be associated with functional transitions within the superfamily and this, in turn, is helpful in predicting the function of previously uncharacterized sequences.

Determining superfamily topology without first knowing the identities of the protein families in the data set is not an easy task. It requires us to calculate boundaries in sequence space based solely on sequence similarity while keeping in mind that the relationships between sequence, structure, and function within a protein superfamily are complex and far from clear. Two closely related superfamily members may share nearly identical sequences, with a few amino acids accounting for the different functions they perform (Seffernick et al., 2001). More divergent families within a superfamily may still share a similar structure in which at least the active site residues associated with the superfamily-common partial reaction are conserved despite sharing a low level of sequence identity (Brenner et al., 1998), (Glasner et al,. 2006). Consequently, we are unable to draw reliable conclusions from local sequence-sequence comparisons. Fortunately, when we aggregate all local sequence comparisons into large-scale protein similarity networks, the results we obtain are much more informative (Enright and Ouzounis, 2000) although they typically lack sufficient resolution to detect topological boundaries between neighboring families. The approach we describe here builds on available similarity network analysis techniques to design a process for identifying topological boundaries in a given superfamily sequence set.

Much of the current research in the field of sequence similarity network analysis has focused on qualitative analysis based on network visualization. Tools such as BioLayout (Enright and Ouzounis, 2001) and CLANS (Fickey and Lupas, 2004) are able to take an all-by-all BLAST-scored (Altschul et al., 1997) network associated with a set of protein sequences and output a visual representation of that network in two-dimensional and three-dimensional space, respectively. They do this by employing the Fruchterman-Reingold force-directed layout algorithm (Fruchterman and Reingold, 1991), which models the network as a physical network in Euclidian space. The algorithm places the nodes in the network into visually discernible clusters whose distance to one another is a function of their connectivity and BLAST scores. These groups might represent subsets within a monofunctional family, or a collection of strongly related families. Individual groups close to each other in Euclidian space may represent functionally separable families that are nonetheless very similar to one another. The groups that are far apart due to little or no direct connectivity (as can be captured using BLAST as a comparison tool) are a result of sequence divergence. As the distance between groups increases, the degree of functional overlap between proteins represented in the network decreases (Adai et al., 2004). By visualizing these spatial properties of a network, we obtain a reasonable global representation of all sequence data within a superfamily.

While network visualization is a useful tool for hypothesis generation, it does not always accurately define a topology between functional classes of proteins within a superfamily. In order to improve the definition of topology, it is first necessary to delineate boundaries between all distinct pairs of neighboring families in a manner that best approximates functional differences. Visualization constrains us to label these boundaries using cluster distributions in two or three-dimensional space. The network itself, however, is a multidimensional object. If we do not know in advance the distribution of network variance across all possible dimensions we run the risk of inferring the incorrect topology based on statistically insignificant distances (Vlachos et al., 2002).

Even if we ignore the issue of dimensionality, we are still unable to accurately determine topology between functionally distinct protein clusters using just the visual representation of a network. As the number of edges between neighboring clusters increases in large networks, the visual representation deteriorates. The clusters are drawn towards one another by the attractive force proportional to the number of connecting edges. Eventually, the proximity between the clusters blurs the spatial border between them and multiple clusters merge into a single large cluster. The user of the network visualization tool is then left with an incomplete representation of the topological relationships between families in the dataset.

Little previous work has been done to address the issue of visual complexity resulting from excess edges in similarity networks. One current approach is to select a threshold and remove all edges with weights below the threshold (Medini et al., 2006). The threshold is manually adjusted until a value is reached that eliminates many redundant edges while maintaining network connectivity. This approach falls short, however, because not all clusters in the network share equal connectivity. When using a threshold suitable for a majority of edges in the network, clusters of outliers connected to the core of the network with very low edge weights may break away, or clusters with multiple poorly weighted connections will disintegrate. It is therefore difficult to maintain network connectivity, which is needed to determine topology, while filtering edges using only a single threshold value.

The goal of the work we describe here is to develop a better filtration approach that maintains network connectivity while highlighting both individual clusters and the topological relationships between them. To do so, we focus on a quantitative analysis of the clusters within similarity networks. The automated clustering of proteins into families based solely on connectivity within protein similarity networks is an expanding area of research. Building on graph theory-based network clustering techniques (Frivolt and Pok, 2006), algorithms such as TribeMCL (Enright et al., 2002) and RANKPROP (Noble et al., 2005) attempt to isolate tightly integrated sets of nodes using criteria such as edge density and edge weights. These algorithms are parameterized to classify protein sequences into unique families using alignment-based protein similarity networks. The clusters they compute are likely to correspond with spatial clusters of nodes that aggregate together in a force-directed layout, but because the clustering is not based on the spatial proximity of nodes, dimensionality is not an issue.

By clustering the nodes in the network, followed by further analysis, we are able to achieve an effective filtration protocol for reducing the number of edges within a network and elucidating the topology in the associated data set. We accomplish this by first clustering the network into sets of tightly connected components. All edges outside the clusters are then removed from the network, leaving isolated clusters. Next, the clusters are reconnected by reinstating a small set of best-scoring edges between nodes in different clusters. Edges added back into the network represent the boundary between pairs of neighboring clusters and define the topological structure of sequences in the dataset. Finally, we visualize the filtered network using a force-directed layout. The layout of the filtered network qualitatively highlights the topology spanning the clusters, allowing for more intuitive hypothesis generation.

## 2 Methods

### 2.1 Outline of the Network Filtration Protocol

Given an all-by-all BLAST-scored protein similarity network, we want to filter it such that individual families within the network fall into obviously distinguishable clusters and that the sequences most optimally connecting the separate clusters are visible within the network. The protocol for accomplishing this can be summarized as follows:

1. Compute an all-by-all protein similarity network using BLAST;
2. Cluster the nodes in the network and remove all edges that do not connect two nodes in the same cluster;

3. Reconnect the clusters using the minimum number of reasonably weighed edges;

4. Visualize the network using a force directed layout algorithm.

*2.1.1 Computing the Similarity Network*

For any input data set, we carry out an automated BLAST search for every sequence in either the NCBI NR database using default parameters, or a custom database built from selected input sequences. Although skewed expectation values result from running a custom BLAST search compared to running a search against the much larger NCBI NR database, this skew is unimportant relative to the topology of the network itself. The BLAST expectation value (e-value) cutoff for each search is set to one in order not to miss possible connections, although this e-value does not represent a statistically significant match.

Each protein is treated as a node in the similarity network. Whenever a BLAST alignment is returned between two proteins in the data set, we connect these proteins with an edge. Each edge is given a weigh equivalent to the –log of the BLAST e-value.

*2.1.2 Clustering the Network*

After computing the similarity network, we carry out clustering using techniques discussed in chapters one and two. We prefilter the network and run MCL in order to cluster the nodes into families. MCL is our algorithm of choice because of its speed and its reliability when a threshold is applied.

*2.1.3 Reconnecting the Clusters*

After we have isolated the clusters, our goal is to reconnect these clusters using a minimal subset of edges from the original all-by-all network. We strive to reconnect clusters by maximizing the connectivity between closely related clusters while minimizing the presence of redundant edges. We accomplish this by computing edges from all possible minimum spanning trees (Prim, 1957) connecting all clusters, using a modified version of Kruskal's algorithm (Kruskal, 1956). These edges, defining the topology between clusters, are added back into the network. Edge weights are rounded to integer values when computing all minimum spanning trees to help address the noisy nature of BLAST e-values. The detailed procedure for our cluster reconnection algorithm is as follows:

1. Create an empty graph list gL and an empty edge list eL. Go to step 2.
2. For each cluster X outputted by tribeMCL, create a graph gX such that all edges from the original unfiltered network connecting the nodes in X are present in gX. Add gX to list gL. Go to step 3.
3. Select all intercluster edges from the unfiltered network that are not present in any graph gX in gL. Add these edges to eL. Go to step 4.
4. Sort edges in eL from largest to smallest edge weight. Go to step 5.
5. If the length of eL is zero, return all nodes and edges present in gL. This is the final filtered network. Otherwise go to step 6.

*2.1.4 Visualizing the Network*

We visualize the final filtered network in Cytoscape (Shannon et al., 2003), an open-source Java-based program originally designed to display protein-protein interaction networks. Cytoscape allows users to assign multiple attributes to the nodes and edges of a given network and then map a set of colors to these attributes. For example, those nodes that represent functionally categorized proteins can be assigned a color based on their family identity. Edges can also be assigned a color based on whether or not they connect nodes from neighboring clusters, as well as on the statistical significance of the corresponding edge weight. The final network is then displayed using Cytoscape's "organic" layout, a force-directed layout algorithm available within the "yfiles" plugin and a standard part of the Cytoscape distribution.

## 2.2 Data Set Selection

*2.2.1 Designing the Protocol*

In order to design our filtration protocol, we used a gold standard collection of manually annotated sequences (Brown et al., 2006) from the enolase (Babbitt et al., 1996) superfamily. We downloaded 681 enolase sequences from the Structure Function Linkage Database (SFLD) (Pegg et al., 2006). We used this dataset for the development of our protocol because it represents a highly divergent superfamily in which families evolve at variable rates. All edge-weights were derived using the NCBI NR database.

*2.2.2 Testing the Significance of the Generated Network Topologies*

It has long been established that evolutionary proximity corresponds to structural and functional similarity (Perutz et al., 1965). Protein families rooted directly from the same branch point in a phylogenetic tree share a higher degree of similarity than families that are not. With this axiom in mind, we decided to compare how network topology relates to evolutionary branching in a well-studied phylogenetic tree. Our goal was not to correlate topology with evolution, but rather to examine the manner in which protein structural and functional similarities could be inferred from a network.

To generate a test dataset, we focused on the kinase superfamily (Manning et al., 2002). In a recent study (Scheeff and Bourne, 2005), the phylogenetic tree for the kinases was generated using rigorous stochastic optimization (Ronquist and Huelsenbeck, 2003) that incorporates both sequence and structural information. The resulting tree encompasses the evolutionary history of 21 kinases, each one from a unique family. The families divide into nine different kinase functional classes. We searched for these families in the KinBase [http://kinase.com/kinbase/] and the KinaseNet [http://www.kinasenet.org] kinase sequence databases. Thirteen of the families were found in one or both of the databases. These families encompassed all nine classes, and encapsulated a total of 527 sequences. We used the sequences to generate a filtered network representation of the kinase superfamily. All edge-weights were derived using a custom database.

# 3 RESULTS

*3.1 Visualizing the Topology in the Enolase Superfamily Network*

We compared the unfiltered enolase superfamily network from our development dataset to the network output by our filtration protocol. Figure 3.1 shows the unfiltered network. Each node is labeled a distinct color based on the SFLD curated family assignment.

The color coding of nodes in Figure 3.1 makes clear the strong presence of family based clusters within the similarity network. The boundaries between the clusters, however, are generally not clear. It is also difficult to see, through a purely qualitative analysis, how the families transition from one cluster to another. Furthermore, while we observe the presence of certain separate clusters due to the color coding based on characterized family assignment, a researcher visualizing a previously uncharacterized superfamily for which high quality annotation is unavailable would likely be unable to distinguish between adjacent components of the network.

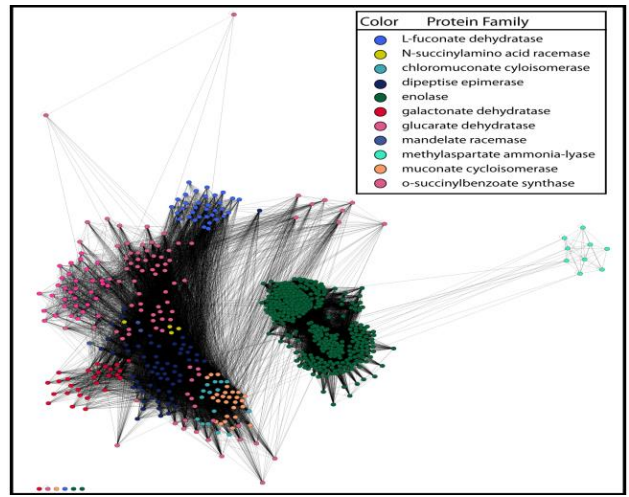

**Fig. 1** Unfiltered Enolase similarity network. Edge-weighed force-directed representation for the pairwise BLAST similarities in the enolase superfamily. Nodes of the same color group together in two-dimensional space, but it is difficult to distinguish which nodes are responsible for the transition between neighboring families.   Certain large spatial clusters are composed of nodes belonging to multiple families. While the nodes in a given family do tend to co-locate, this is only discernible due to their shared color scheme.  This would not have been visible if the identities of these families were not known prior to generating the network.

______________________________________________________________________________

The corresponding filtered network is shown in Figure 3.1 and demonstrates the final output of our network filtration protocol. The enolase superfamily has been separated into clearly distinguishable components by our protocol, corresponding, for the most part, to known protein families. These components are connected by edges that designate pairs of components as neighbors. Edge color defines how closely the components neighbor one another, with the least significant edges shown in blue. The topological relationships between components are easy to detect by direct inspection of the network layout.

The overall connectivity of the family relationships within the enolase superfamily fail in some cases to reflect relationships inferred from highly curated observations derived using experimental methods. For example, the blue edges shown in Figure 3 connecting the OSBS/NSAR and the enolase [family] cluster reflect e-values that range from $10^{-29}$ - $10^{-74}$. Because they are both of low statistical significance and highly complex, the most difficult of the family relationships to capture for this superfamily are those relating the families in the muconate lactizing enzyme (MLE) subgroup.

## 3.2 Structure, Function, Topology and Evolution in the Muconate Lactonizing Enzyme Subgroup

*3.2.1 Introduction to the Muconate Lactonizing Enzyme Subgroup*

The MLE subgroup is a well-studied subset of the enolase superfamily (Glasner et al., 2006). Our enolase dataset represents six catalytic reactions from the MLE subgroup. These include muconate cycloisomerase (MLE I), choloromuconate cycloisomerase (MLE II), Dipeptide epimerase (DipEp), N-succinylamino acid racemase (NSAR), and o-succinylbenzoate synthase (OSBS). Proteins in the OSBS family are particularly difficult to classify because they are highly divergent. Some members share less than 15% pairwise sequence identity with other members of the family. Additionally, certain OSBS enzymes are capable of catalyzing both OSBS and NSAR reactions (Palmer et al., 1999; Sakai et al., 2006). Despite this divergence and promiscuity, careful phylogenetic analysis has revealed that members of the OSBS family (including the OSBS/NSAR enzymes) are monophyletic and more closely related to one another than they are to other families in MLE subgroup (Glasner et al., 2006).

We wanted to explore how this messy interplay of sequence, structure, function, and evolution within the MLE subgroup correlates with network topology. We therefore examined in more detail the topology of the subgraph in the filtered enolase network corresponding to the MLE subgroup (Figure 3.2B).

*3.2.2 Clustering the MLE Subgroup*

The MLE I and MLE II families, which catalyze very similar isomerization reactions, group together in a single cluster. The DipEp family is split across four clusters of sizes one, 11, 15, and 32, respectively. As expected, the divergent OSBS family was distributed across multiple clusters of various sizes. Seven clusters were composed of only a single node. Five clusters each contained between three and eight nodes. The remaining three clusters contained between 12 and 27 nodes.

One of the OSBS clusters includes several proteins annotated as NSARs in the SFLD. Three of these have been experimentally characterized and are promiscuous for both OSBS and NSAR activities (Sakai et al., 2006). The functions of the other proteins annotated as NSAR or OSBS in this cluster have not been experimentally determined, but phylogeny and comparative genomics suggest that while some are physiologically required for OSBS activity, others are more likely to function as NSARs in the cell (Glasner et al., 2006).

*3.2.3 OSBS Connectivity*

The OSBS family forms a monophyletic group in the MLE subgroup phylogeny (Glasner et al., 2006). We therefore had expected there to be a direct path connecting all OSBS clusters. For the just the families from the MLE subgroup. All other families have been deleted from the layout depicted in 4A. The OSBS proteins in the OSBS/NSAR cluster do not directly connect to other members of the OSBS family, despite being more closely related to the OSBS family than are other families within the superfamily. Interestingly, structural superposition shows that the structurally characterized OSBS/NSAR from Amycolatopsis is more similar to an MLE (lower RMSD) than to other structurally characterized OSBSs (Glasner, 2006).

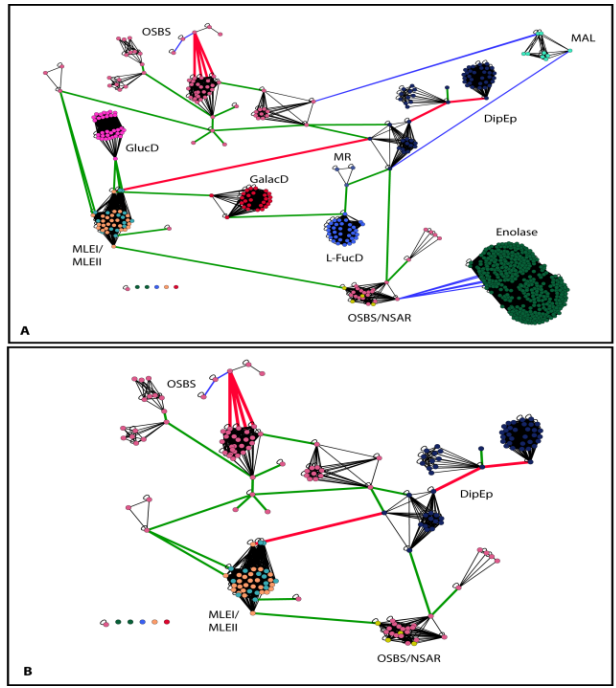

**Fig. 2** Enolase similarity network. (a) Unweighted force-directed layout representations of the enolase similarity network after processing with our filtration protocol. Edges between nodes in the same cluster are colored black. Edges connecting nodes from neighboring clusters are colored blue, red, and green, based on edge weight. Blue edges have an edge weight of less than 10 (e-value > 1 x 10-10). Green edges have an edge weight between 10 and the prefiltering threshold (33, corresponding to an e-value = 1 x 10-33). Red edges have an edge weight greater than the threshold (e-value < 1 x 10-33). Parts of the network have been positioned manually to minimize overlap between red edges. Nodes clustered into the same functional class are clearly visible as discrete circular clusters within the network. Many of these clusters are highly homogeneous with respect to the color assignments generated from SFLD annotation (the names of the protein families have been added by hand for this figure). The global topology of the clusters is easy to distinguish. (b) Subgraph of the enolase network in Figure 3.2 containing

_________________________________________________________________________________

most part this was the case. Eleven of the clusters were connected by a direct path, uninterrupted by the presence of sequence from other families. Edge weights bridging the gap between these clusters ranged from six to 36. One of the 11 clusters connects to AEE with an edge weight of 18. Another connects to MLE I/II with an edge weight of 13.

Despite the connectivity between most OSBS clusters, the OSBS/NSAR cluster does not directly connect to the other 11 OSBS clusters. Instead, it connects to both AEE and MLE I/II with edge weights of 29. This was quite unexpected. The OSBS proteins in that cluster appear closer in sequence space to members of other families then they do to the members of the family with which they share the same function.

There is little evidence to explain this discrepancy except to note that BLAST e-value is not a good enough metric to resolve this type of complexity. While the topology of a BLAST-based filtered similarity network is useful as a hypothesis generator, we are unable to use that topology in order to draw a definitive conclusion.

*3.2.4 Interpreting the Significance of Neighboring Clusters in a Filtered Similarity Network*

Our investigation of the MLE subgroup revealed that the presence of an edge between two distinct clusters is not necessarily a good indication of evolutionary proximity. Rather, the edge implies that the proteins in the two neighboring clusters share some degree of similarity as it can be identified by the comparison method used, in our case the BLAST algorithm, which in turn implies that the proteins share some degree of functional similarity. For the network shown in Figure 3.2B, all clusters in the group have already been validated as sharing some degree of functional similarity by definition—they are all members of the enolase superfamily, each protein of which performs a common partial reaction mediated by a conserved constellation of active site residues that in most cases are easily identified by BLAST (Babbitt, 1996).

It is important to emphasize that we currently have no way of inferring the degree of similarity between two neighboring clusters. Any conclusions we draw about the similarity between two clusters connected by an edge can only be made relative to all other nodes that these clusters do not neighbor. For example, the 375 member enolase family is a direct neighbor to the cluster of OSBS/NSAR sequences. The edge weight connecting the two clusters is zero, indicating that the BLAST alignment between enolase and OSBS/NSAR is not statistically significant. Based on this data we are unable to interpret how much functional similarity is shared between the enolase and OSBS/NSAR clusters. We can, however, hypothesize that because the enolase family has no other neighbors in the network, the degree of overlap between enolase and all other proteins in the dataset is no more significant than the degree of overlap between enolase and OSBS/NSAR. As illustrated by this example, when drawing a hypothesis from a given network topology, it is important to consider not only all pair-wise neighbors, but also the set of all pair-wise clusters that do not neighbor one another.

## 3.3 Examining Protein Kinase Network Topology

*3.1 Summary of the Kinase Network Topology*

We generated an all-by-all kinase similarity network (Figure 3.3A), which we then filtered using our protocol (Figure 3.3B) to produce 20 individual clusters. Nineteen pairs of neighboring clusters define the topology, indicating that no cycles are present. Three of the clusters are composed of multiple families belonging to the same functional class. Functional classes in the kinase superfamily designate groups of evolutionary related families frequently subject to similar functional regulation within the cell (Hanks and Hunter, 1995). Twelve clusters encompass all sequences from a single family within the dataset, while the five remaining clusters each contain a subset of sequences from a unique family.

Eight of the nine functional classes are well connected (Figure 3.3B). Any non-cyclic path between two members of a single well connected functional class contains only sequences from that particular functional class. This does not apply to the atypical kinases (AKs). No atypical kinase family connects directly to a second atypical kinase family.

Cluster degree, defined as the number of neighbors to a given cluster, is not uniform across the network. One cluster has degree eight, one cluster has degree four, three clusters have degree three, two clusters have degree two, and thirteen clusters have degree one. Cluster hubs, which neighbor multiple clusters at the same time, are clearly distinguishable in the network. This clustering information can be ascertained directly just by looking at the final network layout.

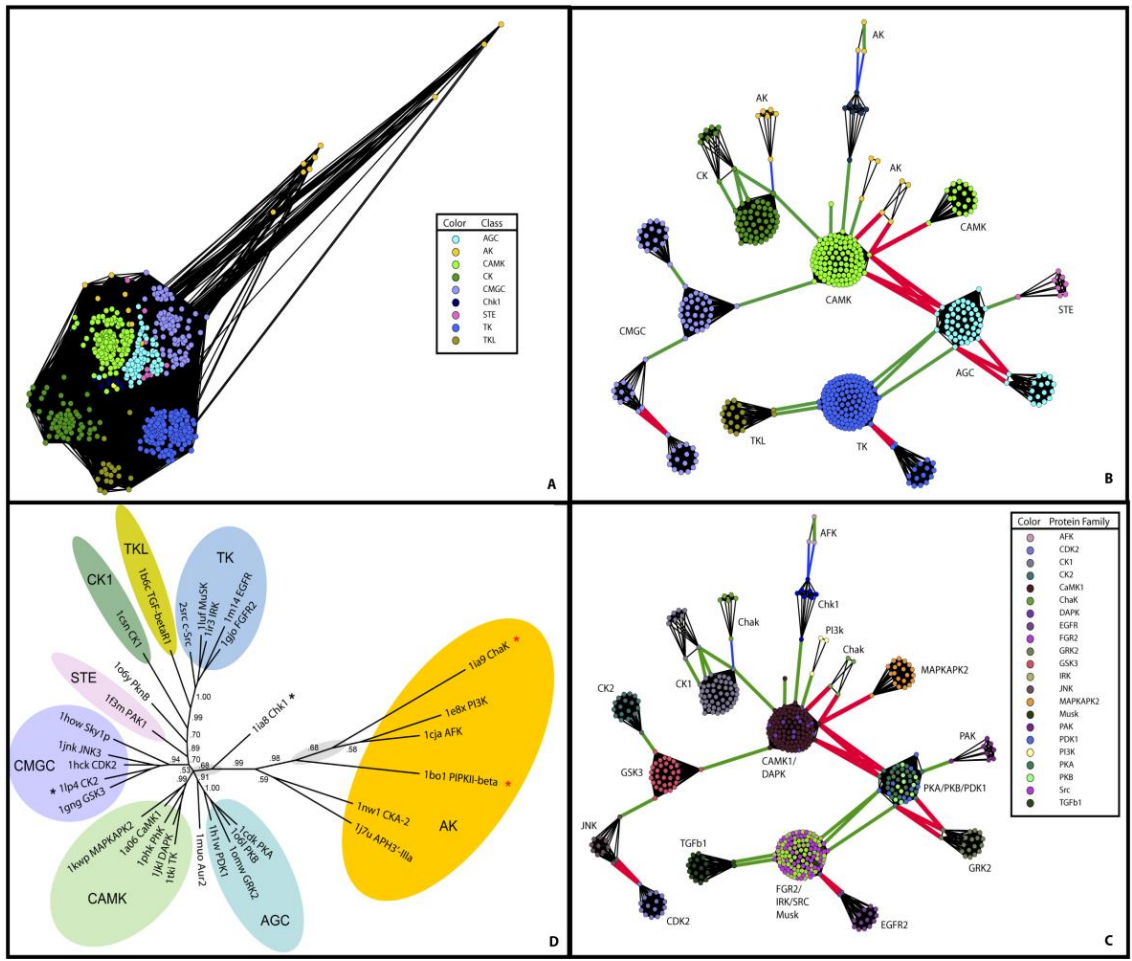

**Fig. 3**. Kinase similarity network. (a) Edge-weighed force-directed representations for the pairwise BLAST similarities in the kinase superfamily. Nodes are colored by functional class and individual functional classes group by color within the network. Node classifications to functional classes and families were obtained from the KinBase and KinaseNet databases. No direct connectivity is discernible from this cluttered network representation. (b) Unweighted force-directed layout representation of the filtered similarity network colored by functional class. Edges connecting nodes from neighboring clusters are colored blue, red, and green, based on edge weight. Blue edges have an edge weight of less than 10. Green edges have an edge

weight between 10 and the prefiltering threshold of 42. Red edges have an edge weight greater than the threshold. Parts of the network have been positioned manually to minimize overlap between intercluster edges. The clusters correspond to either individual families or individual functional classes. Eight of the nine functional classes are well connected. According to the topology, the CAMK functional class is a central hub in the network, connecting five of the functional classes. (c) Unweighted force-directed layout representation of the filtered similarity network colored by family. (d) The kinase superfamily phylogenetic tree, optimized with Mr. Bayes using both sequence and structural data. Both families and functional classes are indicated in the tree. Leaves in the tree correspond to individual families. The labeled ovals encompassing multiple families, correspond to functional classes, as defined by (Scheeff and Bourne, 2005). Each oval, signifying a unique functional class, is labeled a unique color. Chk1 is the closest of the typical kinases to the AK functional class. Kinases labeled with a black asterisk are classified differently in the tree compared with the classification produced by Manning (Manning et al., 2002). Reprinted with permission of PLoS Computational Biology from (Scheeff and Bourne, 2005).

______________________________________________________________________________

*3.3.2 Comparing Network Topology to Phylogenetic Branching*

We analyzed branching in the evolutionary tree from the 2005 Scheeff and Bourne study (Scheeff and Bourne, 2005) (Figure 3.3D). Three pairs of functional classes connect directly to a single internal node, while nine pairs of families also descend directly from a single branch point. Seven of these pairs are present in our data set. We are therefore able to compare network topology with phylogenetic branching in 10 functional classes and family pairs by measuring the hop distance between each of the pairs across the network. We defined hop distance as the

___________________________________________________________________

**Table 1.** Comparing Phylogenetic Divergence to Filtered Network Topology Data in Kinase Superfamily

| Kinases With Direct Common Ancestor | Classification | In Same Cluster | Neighbors | Hop Distance |
|---|---|---|---|---|
| TK – TKL | Class | No | Yes | 1 |
| CAMK – AGC | Class | No | Yes | 1 |
| Chk1 – AK | Class | No | Yes | 1 |
| EGFR – FGFR2 | Family | No | Yes | 1 |
| Musk – IRK | Family | Yes | No | 0 |
| CAMK1 – MAPKAPK2 | Family | No | Yes | 1 |
| PKB – PKA | Family | Yes | No | 0 |
| JKN3 – CDK2 | Family | No | Yes | 1 |
| CK2 – GSK3 | Family | No | Yes | 1 |
| AFK – PI3K | Family | No | No | 3 |

Column 1 contains pairs of families and functional classes that are believed to have evolved directly from the same common ancestor. Column 2 specifies whether the kinase pairs are classified as families or functional classes. Column 3 specifies whether or not the kinases appear in the same cluster. Column 4 specifies whether or not the kinases appear in neighboring clusters. Column 5 specifies the hop distance between the kinases, which we define as the minimum number of clusters that must be traversed across the filtered network to connect a given kinase pair. The average hop distance is 1.0.

_______________________________________________________________________________

minimum number of cross-cluster traversals that separate two distinct kinase groups. A hop distance of zero indicates that two groups are in the same cluster. A hop distance of one indicates that the two groups are found in adjacent clusters that neighbor one another. The hop distances between all ten pairs of kinase groups are listed in Table 3.1. The average hop distance and the median hop distance for the 10 pairs are both equal to one. In contrast, the mean hop distances between all pairs of functional classes and all pairs of families, are 2.22 and 2.51 respectively. These results informally imply that for this system, protein functional groups evolving directly from a single ancestor have a greater propensity to neighbor each other or cluster together in the filtered network. In other words, if two functional groups are not neighbors then they are less likely to have evolved directly from a single ancestor.

Eight of the 19 neighboring cluster pairs corresponding to seven pairs of functional classes were of indeterminate significance. These indeterminate pairs (TK-AGC, STE-AGC, CAMK-AK, CK-AK, CMGC-CAMK, Chk1-CAMK, CK-CAMK) consisted of neighboring clusters from distinct functional classes that had not diverged directly from a single common ancestor. The significance of these pairs is not known at this time. We are, however, able to state that for our BLAST-based network, over half of the neighboring clusters in the filtered kinase network are consistent with known evolutionary relationships.

*3.3.3 Determining the Nearest Neighbor to the Atypical Kinases*

The proteins in the atypical kinase class differ from other members of the kinase superfamily in that they do not share certain sequence and structural motifs common to all typical kinase proteins. One of the goals of the Scheeff and Bourne study was to determine which kinase class had the greatest evolutionary proximity to the atypical kinases. According to the phylogenetic tree, the AK class and the channel kinase (Chk1) class directly evolved from the same common ancestor. However, the bootstrap value connecting AK and Chk1 to an internal branch point was not reliable enough for the authors to draw a definitive conclusion. Furthermore, a second phylogenetic tree stochastically optimized using just sequence data showed that the choline kinase (CK) class, rather than the Chk1 class, connected to the atypical kinases, albeit again at a very low bootstrap value. The authors presented arguments demonstrating that both CK and Chk1 make good candidates as the closest evolutionary link to the atypical kinases, and that one or the other is the actual link.

In the filtered network representation of the kinase superfamily (Figure 3.3C), both CK and Chk1 connect to members of the AK class. Chk1 neighbors the actin-fragmin kinase (AFK) family, while CK neighbors a subset of the channel kinase (Chak) family. The remaining atypical kinase sequences, which include the phosphoinositide 3-kinase (PI3K) family and a subset of the Chak family, connect to the calcium/calmodulin-dependent kinase (CAMK) class, which connects directly to both CK and Chk1 in the network. Although the BLAST e-values underlying these results are not statistically significant, the results themselves are consistent with the two candidates for nearest evolutionary neighbor derived using phylogenetic analysis.

# 4 DISCUSSION

## 4.1 Network Topology as a Metric of Functional Similarity

Our results indicate that protein families which are not neighbors in the kinase network are less likely to descend directly from the same common ancestor. Since evolutionary distance reflects structural and functional proximity, these results suggest that a filtered network topology may be useful for developing hypotheses about structural and functional similarity. Individual clusters within a filtered network correspond to whole families or sets of functionally similar families within a superfamily. The topology between these clusters suggests the degree of functional similarity between distinct families and functional classes. Families that do not neighbor one another are less likely to share structural and functional properties than neighbors within the network.

These properties suggest that filtered similarity networks are a useful tool for discriminating sequence clusters in order to provide a starting point for predicting functional relationships and properties in poorly understood protein data sets. A researcher examining a large superfamily with few functionally characterized members will be able to apply our protocol and generate a simple visual representation of all sequences in the data set. Upon visual inspection it should be clear which uncharacterized proteins group together with members of known families. These proteins are likely to be functionally similar to the families with which they cluster, influencing the scope of the experimental assays necessary to characterize function. Additionally, certain clusters will be composed entirely of uncharacterized sequences, indicating the presence of new families. The characterized properties of clusters neighboring unknown families could help constrain the possible functions of these uncharacterized sequences. The network topology will influence hypothesis generation, which in turn allows the researcher to prioritize functional assays in order to efficiently characterize new functions within a superfamily.

Based on the intuitive nature of the filtered network layout, it is possible to investigate functional properties just by visual inspection of the network. However, unlike in the all-by-all network view, the topological boundaries between clusters in the filtered network are clearly defined prior to visualization in two- or three-dimensional space. Our filtration protocol allows researchers to automate the process of network generation relevant to function prediction, without relying on Euclidian distances across dimensionally reduced spatial representations of large multidimensional sequence datasets.

## 4.2 Contrasting Network Analysis with Phylogenetic Analysis

Using our protocol, we are able to suggest relationships between typical and atypical kinases that have previously required combining data from two separate phylogenetic trees. At the same time, we are unable to recapitulate the conclusion that atypical kinase families interconnect to form the AK functional class. Clearly, a similarity network topology does not hold the same statistical significance as a stochastically optimized phylogenetic tree. It is, however, possible to foresee research problems that lend themselves better to network analysis then to phylogenetic analysis.

Filtered homology networks are not as rigorous as phylogenetic trees in representing sequence relationships. The topology of phylogenetic trees is based on detailed mathematical models of protein evolution (Cavalli-Sforaza and Edwards, 1967). In contrast, our protocol uses a heuristic approach that elucidates structural and functional similarity from global sequence comparisons without being restricted by any one model. This heuristic approach provides a useful additional tool for researchers seeking to extract potentially important features within a large sequence data set. A few minutes of computation time is all that is required to filter and visualize a similarity network based on several thousand sequences. The process is entirely automated, requiring no a priori assumptions about the functional identity of the sequences within the network. By contrast, an optimal phylogenetic tree can only be computed using a limited subset of sequences in a multiple sequence alignment due to the computational complexity of properly aligning a large and diverse set of sequences. In a large data set, the subset of sequences in the multiple sequence alignment captures only a small fraction of the total available information. Furthermore, selecting the best multiple sequence alignment subsets is a subjective task for the researcher, leading to the risk of bias in the results derived from the data. Even when a well-prepared data set is ready for phylogenetic analysis, evaluating the optimal evolutionary tree usually takes hours of computation time (Laget and Simon, 1999). Therefore, a filtered similarity network serves as a good substitute to a phylogentic tree in those cases when rapid hypothesis generation across a large, diverse dataset takes priority over rigorous statistical significance.

It is also worth emphasizing that the network topology representation of a sequence data set includes connectivity properties not accessible through a dendrogram or phylogenetic tree. As shown in the kinase network, the degree of connectivity varies from cluster to cluster. This variability allows us to distinguish CAMK as a major hub in the network, which connects five distinct functional classes. This is not at all clear from the phylogenetic representation, where individual proteins connect indirectly through pathways of interior nodes.

The significance of similarity network hubs is unknown at this time. We hypothesize that such hubs may serve as indicators of proximity to phylogenetic branch points. Future studies will test these and other hypotheses in order to determine if the presence of hubs signifies evolutionary relationships that are not discernible within a phylogenetic tree. Filtered network topologies provide context and terminology that make it possible to examine the importance of hubs in more detail.

## 4.3 Caveats

In our current implementation we are unable to display inter-family relationships with complete accuracy, partially because our greedy approach to reconnecting the clusters relies solely on a simple scoring metric for edges (e-value) of limited precision. Local alignments made by BLAST across pairs of motifs with significantly different lengths may lead to misleading connections within a similarity network. More sophisticated similarity comparisons, such as profile-profile alignments and hidden Markov models, could lead to more accurate network topologies.

We are also aware that the quality of a network topology depends on how well the functional groups in the dataset separate out in the first place. Kinase functional groups are more discrete with respect to one another then families in the MLE subgroup, for example. This in turn leads to better clustering, and a more meaningful topology. Knowing in advance the separability of functionally similar groups in a network would give us some measure of topological reliability. Currently, we are unable to infer the discreteness of network components from only sequence data.

# 5 CONCLUSIONS

We have developed a protocol for filtering protein superfamily similarity networks. The protocol divides an input network into discrete components while at the same time emphasizing the topology that best connects the components together. We have shown that individual clusters in the filtered networks correspond to families and classes of functionally similar proteins.

Additionally, we provide evidence that neighboring clusters represent more similar sets of proteins than clusters which are distant. Our results suggest that network topologies in a protein similarity graph, as defined by our filtration protocol, embody a meaningful representation of structural and functional similarities between individual functional groups within a protein superfamily.

In addition to defining topology, our filtration protocol also leads to a more meaningful visualization of the data within the network. An unfiltered network resembles a "hairball," where clusters are often difficult to distinguish from one another and overlapping edges make it difficult to see significant connections. By filtering the network prior to visualization using a force-directed layout algorithm, we are able to directly count the number of clusters and see precisely how these clusters connect to one another. This direct global view provides a useful alternative for summarizing a large data set in a single easy-to-comprehend image. Our protocol can be used to output a simple representation of otherwise complex information, thereby facilitating the generation of useful hypotheses relevant to the data set in question.

The use of global protein similarity networks in the bioinformatics research community continues to rise. Our filtration protocol builds on existing network techniques to yield a comprehensive understanding of protein superfamily data. We believe the protocol serves as a foundation for developing new techniques capable of making meaningful structural and functional predictions based only on sequence information.